# Room temperature self-assembly of mixed nanoparticles into complex material systems and devices

Masood Naqshbandi[1,2], John Canning[1,2*], Brant C. Gibson[3] & Maxwell J. Crossley[2]

[1] interdisciplinary Photonics Laboratories (*i*PL), School of Chemistry, The University of Sydney, NSW 2006, Australia

[2] School of Chemistry, The University of Sydney, NSW 2006, Australia

[3] School of Physics, The University of Melbourne, Parkville, 3010, Australia

Contact:

John Canning

iPL, Room 222 Madsen Building F09,
The University of Sydney, NSW 2006, Australia

T: +61 2 93511934   F: +61 2 93511911

john.canning@sydney.edu.au

**Room temperature self-assembly of mixed nanoparticles into complex material systems and devices**

The ability to manufacture nanomaterials with complex and structured composition using otherwise incompatible materials increasingly underpins the next generation of technologies. This is translating into growing efforts integrating a wider range of materials onto key technology platforms[1] – in photonics, one such platform is silica, a passive, low loss and robust medium crucial for efficient optical transport[2]. Active functionalisation, either through added gain or nonlinearity, is mostly possible through the integration of active materials[3, 4]. The high temperatures used in manufacturing of silica waveguides, unfortunately, make it impossible to presently integrate many organic and inorganic species critical to achieving this extended functionality. Here, we demonstrate the fabrication of novel waveguides and devices made up of complex silica based materials using the self-assembly of nanoparticles. In particular, the room temperature fabrication of silica microwires integrated with organic dyes (Rhodamine B) and single photon emitting nanodiamonds is presented.

Rhodamine B is used to help characterise the attenuation and transmission properties of the waveguides whilst the nitrogen vacancy (NV) doped nanodiamonds[5] allow for the demonstration of robust, room temperature, single photon emitting sources integrated into silica. Such single photon sources have been studied extensively since their discovery[6] and have been used for many applications[7-11] including quantum cryptography for secure communications and computing[12]. However, whilst they are sufficiently robust to integrate into materials such as soft glasses (having been drawn, for example, into tellurite optical fibres at 700 °C[13]) the problem of integration into viable transmission platforms such as silica has not been resolved. The work described here, by contrast, lays the foundation for a new approach using room temperature self-assembly that can enable ubiquitous future quantum communications, optical computing and more based on silica.

Silica is a key platform material, most commonly manifested as photonic waveguides including micro- and nano-wires. Interest in such photonic wires is increasing, driven by a need for more compact and low loss structures in a broad range of applications spanning sensing[14], micro- and nano- photonics[15] through to optical interconnects for faster and reduced thermal footprint computing[16]. These wires are essentially waveguides with an air cladding, generating a high numerical aperture and permitting easy access to a significant evanescent field which is ideal for sensing and diagnostic applications as well as for optical components. Generally, different bottom-up approaches have been taken to manufacturing micro- and nano- wires[17-20] using a variety of materials such as silica, silicon, phosphate and various polymers[21]. For silica the most successful approach has been largely top-down tapering of optical fibres using $CO_2$ lasers[22] and naked flames[23]. Of these methods, nanowires drawn using a naked flame have produced the lowest transmission loss waveguides, $\alpha < 0.1$ dB/cm. However a problem with these methods is the need for high temperatures ($T > 1000$ °C) which automatically prohibits the introduction of numerous species, such as organic dopants.

In this work, a novel bottom-up approach based on evaporative self-assembly of nanoparticles at room temperature to overcome these limitations is used to demonstrate a new pathway to controlling and changing the composition of the waveguide for any desired application. At room temperature, inhibitive processes ordinarily present, such as evaporation and ablation, are circumvented by relying on van der Waals forces and lowest free energy packing configurations alone. Our approach makes it possible to include an array of chemical species into silica microwires, overcoming the limitation of previous 'hot' methods thus greatly expanding their possible uses in a range of different fields.

The silica nanoparticles ($n \sim 1.475$, $\varphi \sim (20\text{-}30)$ nm) are suspended in a polar solvent, water ($H_2O$), with trace amounts of $NH_4^+$ to overcome van der Waals attractive forces and prevent aggregation. Dopants, such as rhodamine B and nanodiamonds ($\varphi \sim (20\text{-}30)$ nm), were introduced into the silica dispersion and mixed thoroughly. Under appropriate conditions, as the solvent is evaporating the nanoparticles are able to aggregate[24] assisted by Brownian motion and through attractive forces slowly pack into a crystal-like configuration of lowest energy. Consequently, drops ($V = 100$ μL) of the silica dispersion were deposited onto a borosilicate glass slide treated with methanol to increase wettability - without this treatment the contact angle is too large preventing wire formation. Instead, the material aggregates into small pieces. The wire formation preferentially occurs in the direction of the evaporating solvent front in combination with aggregation and packing, which leads to the formation of rectangular waveguides through Marangoni-type stresses[25]. As they dry, through surface and internal tensions, they rise up detaching from the substrate - Figure 1a summarises this process. Typical silica microwires that have been fabricated are rectangular (6 μm x 4 μm, chosen to match the index and modal profile of SMF-28 telecommunications fibre) in cross-section and over centimeters in length, producing ultra high aspect ratios ($>10^3$). The microwires are mechanically robust allowing for easy handling and manipulation onto a customized optical waveguide characterisation setup using two optical fibre micropositioners. They can also be easily cleaved, similar to optical fibres, using a ceramic ($Al_2O_3/SiO_2$) tile to introduce a stress fracture. Figure 2a shows a SEM image of the wire and its rectangular cross-section. Figure 1b shows the wire under applied tension prior to manual cleaving. Figure 2b and 2c show close-ups of the wire surface revealing the short range hexagonally close packed lattice structure which is the most efficient packing configuration for uniform hard spheres ($\eta \sim 0.74\%$). The formation of structural defects and nanocavities are minimised and this packing may offer a novel test-bed approach to exploring topological theories of crystal and glass formation.

The evaporation of a single drop of solution is dependant on many factors including both the diffusion of molecules to air from the liquid-vapor boundary and molecules from inside of the drop to the liquid-vapor boundary[26]. As the drop evaporates, its volume shrinks closely described by:

$$\frac{dV}{dt} = -\frac{\pi D(T) M}{\rho RT} W \frac{(1-\cos\theta)}{\sin\theta}(P_S - P_\infty) \qquad (1)$$

where $D(T)$ is the diffusion coefficient, $M$ is the molecular weight, $T$ is the temperature, $R$ is the gas constant, $W$ is the contact base diameter, $\theta$ is the contact angle, $\rho$ is the density and $P_S$ and $P_\infty$ are the vapour pressures of the liquid on the drop surface and of the surrounding atmosphere respectively. From this expression, the rate can be affected by varying the contact angle ($\theta$), adjusting the partial pressure, the concentration, density or changing the temperature, noting that many of the parameters are both temperature and time variant. Anything which slows the evaporation rate will improve wire formation by providing more time for aggregation and packing. We observed, for example, that the smaller the contact angle the longer the wires. In addition to controlling the drop size and evaporation rate, by varying the concentration of nanoparticles, tens to hundreds of microwires up to a few centimeters in length have been produced from a single drop. Low-cost mass production using inkjet deposition is possible. The cross-sectional dimensions also vary with concentration as well as with substrate wettability - the smaller the contact angle the longer and thinner the wires. By varying all these parameters, microwires with cross-sectional widths between 2 and 90 μm were fabricated. This flexible window allows for mode matching to a range of optical fibres and waveguides.

Absorption and fluorescence measurements of the wires self-assembled from combined silica and rhodamine B solutions characterised the wires, demonstrating the integration of volatile organics. Figure 1d shows the coupling of white light through a curved, rhodamine B doped microwire where the green dye absorption leads to a red coloring towards the output end. Similarly, Figure 1e shows coupling of the green HeNe laser ($\lambda = 543$ nm) which excites rhodamine B ($\lambda_{max} = 552$ nm) generating strong fluorescence in the red. The self-assembled rhodamine B doped silica microwire has an estimated propagation loss of $\alpha \sim 3.4$ dB/mm at $\lambda = 1550$ nm for a wire with varying width (input end: 30 μm x 20 μm; output end: 60 μm x 20 μm) and for a radius of curvature, $R_{curv} = 4.43$ mm. The total insertion loss ($\alpha < 30$ dB, $\lambda > 600$ nm) arises principally from mode mismatch coupling loss and attenuation on the glass slide. Mitsui *et al.* have reported formation of colloidal wires via a templating method using polystyrene spheres with $\alpha \sim 2.0$ dB/10 μm[20] and Barrelet *et al.* have reported $\alpha \sim 1.0$ dB/10 μm in CdS nanowires[16]. Tong *et el.* have reported $\alpha < 0.1$ dB/cm at $\lambda = 633$ and 1550 nm[23] for their tapered silica microwire which is single mode with adiabatic input and output coupling. In these experiments, our self-assembled silica microwires are observed to be multimode (Figure 4) with a calculated $V > 132$ assuming a rotationally invariant analog microwire with radius $w = 30$ μm at $\lambda = 1550$ nm. For optical sensing work, multimode propagation is often used as a means of increasing the waveguide collection angle (or numerical aperture), and therefore the signal-to-noise ratio, compared to single mode waveguides. An estimate of the refractive index of the silica microwire is $n \sim 1.475$, similar to that of individual silica nanoparticles as reported, for example by Khlebtsov *et al.*[27]. This value is also close to both fused silica and quartz ($n \sim 1.5$), reflecting tight packing of the nanoparticles. Although the losses of the different silica microwires cannot be directly compared because of the different preparation and measurement techniques, we can comment that the microwire reported here is the first such self-assembled wire. Its total transmission loss, which is undoubtedly a level well above what can be achieved with improved waveguide matching and coupling, lies between that of the top-down fabricated silica optical fibre tapers and that of other templated colloidal microwires made from semiconductor materials. Our losses should be able to approach the best adiabatic tapers produced by top-down, high temperature drawing since there is no intrinsic material loss involved with the insertion loss.

Commercially purchased NV defect containing nanodiamonds were also integrated into the silica microwires as potential single photon sources. Typical particle sizes were $\varphi \sim 50$ nm, almost twice that of the silica nanoparticles. These were readily dispersed amongst the silica particles in water. Despite the size difference and some possible lattice strains, we were able to integrate these into the wires during self-assembly and characterised them using scanning confocal microscopy at room temperature. Figure 5a shows a 40 μm × 40 μm region of a nanodiamond embedded silica microwire; the inset zooms in on a 5 μm × 5 μm region containing a single photon emitting luminescent NV centre. The photon statistics of this centre are shown in Figure 5a, measured at room temperature ($\lambda_{ex} = 532$ nm, 0.5 mW); the background corrected correlation function[28], $g^2$ (t), for this emitter, is shown in Figure

5b. The lowest value, $g^2(0) \sim 0.1$, demonstrates a single photon emitter was addressed. The measured single photon emission rate on one detector was ~30k counts/s under the same excitation conditions. This reflects an effective lifetime of $t \sim$ 33ms, significantly larger than the reported NV defect lifetyime of 17ns[5, 29] suggesting a possible resonance lifetime within the nanodiamond spherical resonator (where $n_d \sim 2.4 > n_{SiO2} \sim 1.5$). The inset of Figure 5b shows the stability over time of the emitter, with no blinking present, an important advantage of these diamond centres over alternative materials.

In conclusion, the self-assembled silica microwires, doped with materials such as rhodamine B and nanodiamonds reported here opens up a new field for compositional control of silica waveguides integrated with organic and inorganic species. Standing to benefit are a host of novel applications spanning optical interconnects, optoelectronics, sensing and diagnostics, plasmonics and metamaterials, quantum computing and communications and more. The low cost, scalable process could, in principle, be adapted to grow directly onto existing technologies for immediate integration including, for example, wires attached to integrated silicon waveguide circuits.

## Methods

**The self-assembly of silica microwires**

The microwires are self-assembled from individual silica nanoparticles with diameter between $\varphi \sim (20\text{-}30)$ nm in an aqueous colloidal dispersion (5 % w/w, pH = 9 with $NH_4^+$ counter-ions) onto a pathology grade borosilicate glass slide. The glass slide was wetted by methanol ($V$ = 100 µL); the silica dispersion ($V$ = 100 µL) was subsequently deposited on top using a hand-held pipette (nozzle $\varphi$ = 2 mm). The solvent is left to evaporate (1 atm, 295 K), which leads to microwire formation.

**Integration of rhodamine B and Nanodiamonds**

Both the organic fluorescent marker - rhodamine B {[9-(2-carboxyphenyl)-6-diethylamino-3-xanthenylidene]diethylammonium chloride [$C_{28}H_{31}ClN_2O_3$]} ($m$ = 0. 1 mg) and the nanodiamonds ($m$ = 0. 1 mg), were introduced in solid form to different silica dispersions (5% w/w, $H_2O$) ($V$ = 2 mL). These dispersions were stirred (10 min) followed by utlrasonication (10 min) to allow mixing. The procedure above for wire formation is then followed.

**Optical characterisation of self-assembled silica wires**

The propagation of light within the silica microwire waveguides was characterised using several different sources: (a) red and green HeNe lasers ($\lambda$ = 632.8 nm, 543 nm) and (b) white light source (1000 W Xe lamp). Input and output light was delivered by standard SMF-28 telecommunications optical fibre whist transmission spectra are collected on an optical spectrum analyser. To obtain absorption and transmission measurements of rhodamine B absorption within the microwire, a butt-coupling set-up with 3-axis micropositioners was used. A curved silica microwire is placed on the edge of a low-index borosilicate glass slide and two SMF-28 fibres guide light into the microwire and collect the light from the output, which is then processed on the optical spectrum analyser (OSA). An optical image of the set-up is shown in Figure 1c.

# Imaging and single photon correlation measurements of NV containing nanodiamonds

A custom built scanning confocal microscope, with a x100 ($NA = 0.95$) air objective and bandpass filter ($\lambda_{em} = 650\text{-}750$ nm), was used to measure NV centre emission from the silica-nanodiamond samples. Photon statistics from each fluorescent region on the confocal images were studied using a Hanbury Brown and Twiss interferometer[30] to identify single NV centres. The photon emission statistics were measured after excitation with a frequency doubled Nd-YAG laser ($\lambda_{ex} = 532$ nm, 0.5 mW). The background corrected correlation function, $g^2(t)$, shown in Figure 5b, was determined using the methods described by Brouri *et al* [28]. The single emitter was found to exhibit a characteristic single photon dip ($< 0.5$) in the second order correlation function, $g^2(0)$, at zero delay time ($t = 0$).

**Acknowledgments** We thank the University of Sydney for a Gritton postgraduate scholarship awarded to M.N. This work was supported by Australian Research Council Discovery Project grants (DP0770692, DP0879465).

**Author contributions** J.C., M.N., B.C.G., and M.J.C. conceived the ideas and designed the experiments. M.N. and B.C.G. carried out the experiments. J.C., M.N., B.C.G. and M.J.C. analysed the data. M.N., J.C., B.C.G. and M.J.C. contributed to writing the manuscript.

**Author Information** The authors declare no competing financial interests. Correspondence should be addressed to J.C. (john.canning@sydney.edu.au).

**Figure captions:**

Figure 1. Self-assembly and light guidance of silica microwires: (a) Evaporative self-assembly formation of silica microwires. Optical images photographed with a microscopic CCD camera after drop deposition ($t$ = 1, 3, 9 and 15 min). As the solvent evaporates, silica nanoparticles aggregate and break into curled microwires;

(b) Optical image of a microwire (*l* ~7 mm, w ~10-20 μm) suspended in air, fixed at both ends; (c) Optical microscope image of a rhodamine B doped microwire in the butt-coupling setup (90º bend); (d) Microwire in *(c)* with propagating white light; and (e) Another rhodamine B doped microwire with propagating green light (HeNe, $\lambda$ = 543 nm).

Figure 2. SEM images of the silica microwires: (a) Self-assembled silica microwire with rectangular cross-sectional dimensions of (25 μm x 10 μm); (b) Close-up of *(a)* showing smooth top surface and slightly rougher break (side) surface; (c) Extreme close-up of *(a)* showing the hexagonally close packed-like structure; (d) A sharp cleave induced by stress fracture applied using a ceramic tile.

Figure 3. Optical transmission and fluorescence measurements: (a) transmission spectrum of white light thorough an undoped silica microwire 'undoped/white light', and through a rhodamine B doped silica microwire 'doped/white light'. A dip in the transmission ($< \lambda$ = 590 nm) is observed due to the absorption peak of the rhodamine B; (b) Fluorescence spectrum recorded using green laser excitation (HeNe, $\lambda$ = 543 nm). Transmission is observed in the 'undoped microwire'; in the rhodamine B 'doped microwire' red fluorescence is observed ($\lambda$ ~ 620 nm).

Figure 4. Modal profile in far field. Image and intensity spectra of a silica microwire (SEM inset) showing the multimode characteristic of a microwire with dimensions of ~ 25 um x10 μm.

Figure 5. (a) Diamond NV− photoluminescence intensity scanning confocal map of 40 μm × 40 μm region of an ND embedded silica microwire sample. The 5 μm × 5 μm zoomed region shows NV − PL intensity from an embedded single photon emitter. Note: counts shown were recorded from a single detector; (b) Background corrected second order autocorrelation function (blue circles) of a single NV − centre embedded in the silica microwire, shown in *(a)*, measured at room temperature. The value of *g*$^2$(0) ~ 0.1 indicates single-photon emission. The solid red line represents a single exponential fit of the photon statistics. The inserted figure shows the photostable emission from the single emitter identified in *(a)*.

Figure 1

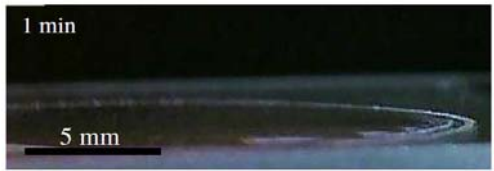 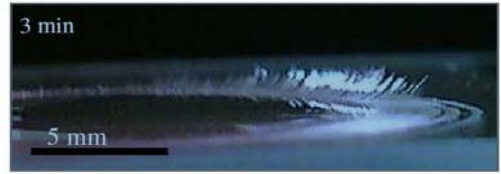
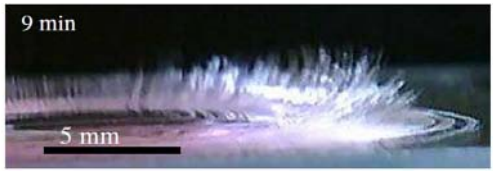 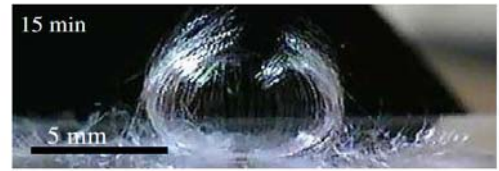
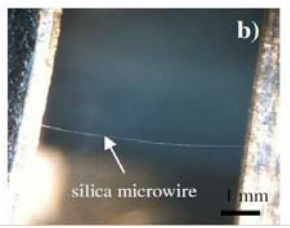 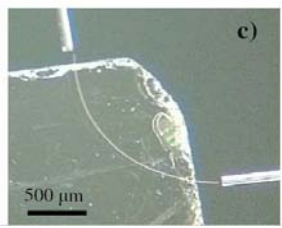 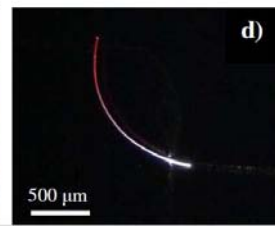 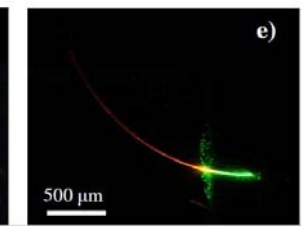

Figure 2

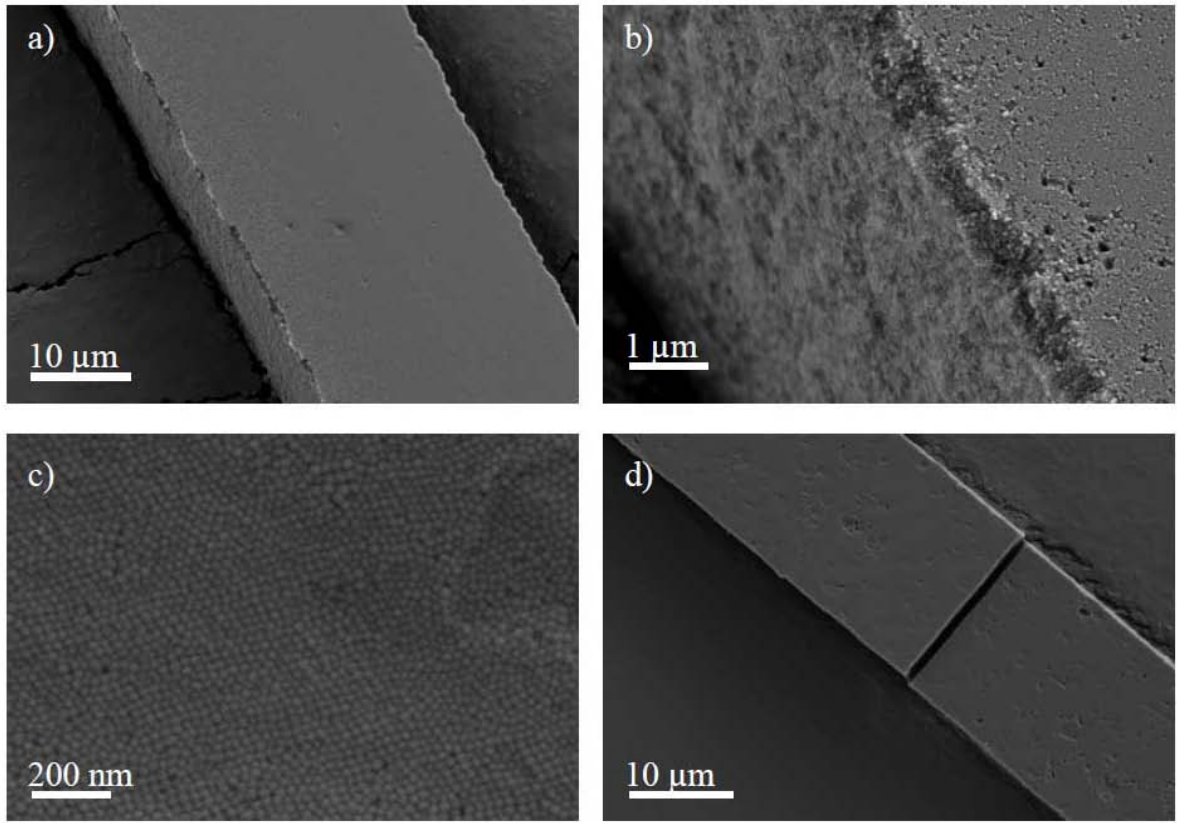

Figure 3.

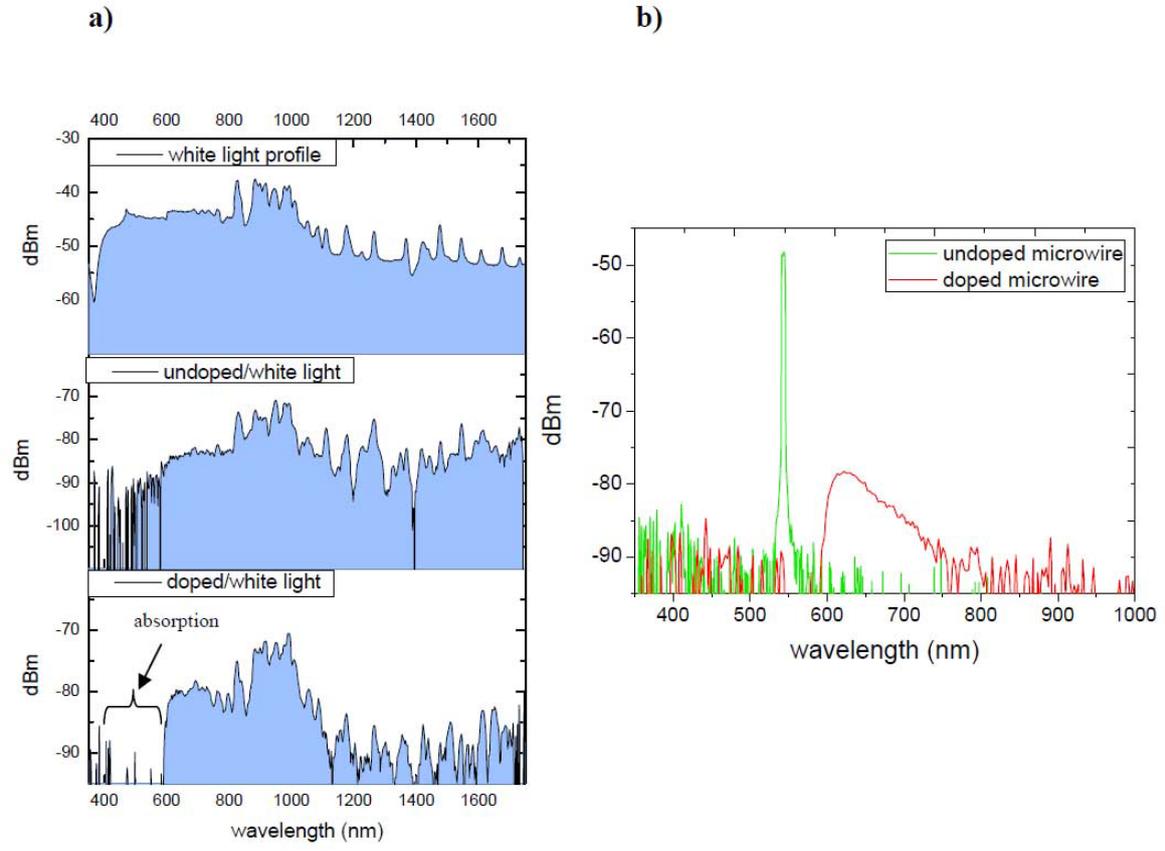

Figure 4.

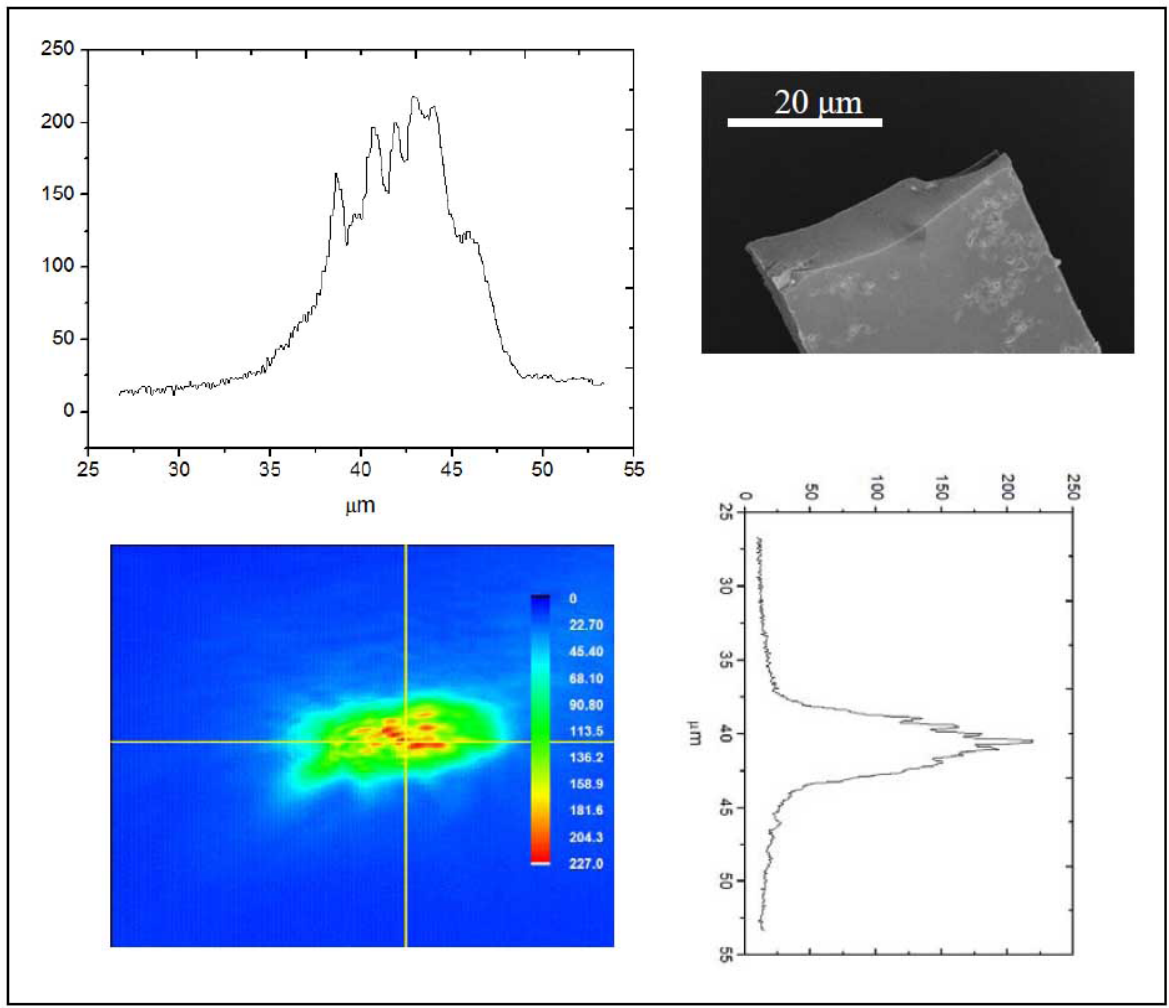

Figure 5.

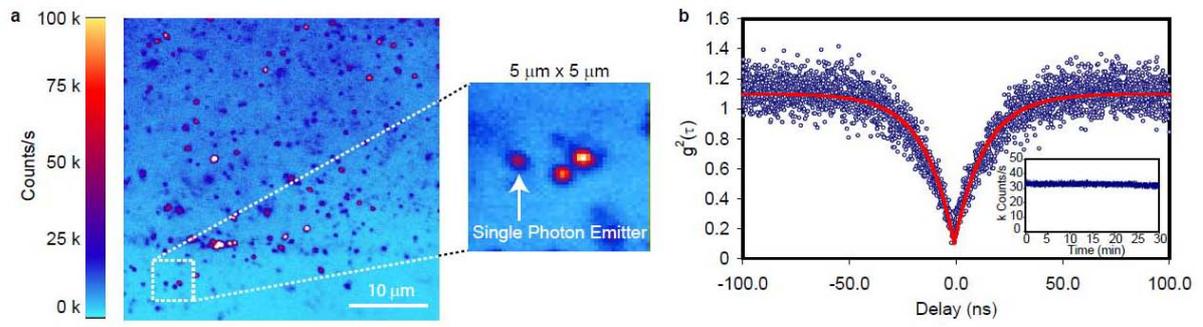